\documentclass[10pt,twocolumn, amssymb, amsmath, aps, superscriptaddress, showpacs, footinbib, prb]{revtex4-1}
\pdfoutput=1
\usepackage{graphicx}

\newcommand{\iso}{\text{iso}}
\newcommand{\sub}{\text{sub}}
\newcommand{\Cels}{$^{\circ}$C}

\begin{document}

\title{Interplay of atomic displacements in the quantum magnet (CuCl)LaNb$_2$O$_7$}

\author{Alexander A. Tsirlin}
\email{altsirlin@gmail.com}
\affiliation{Max Planck Institute for Chemical Physics of Solids, N\"{o}thnitzer
Str. 40, 01187 Dresden, Germany}
\author{Artem M. Abakumov}
\email{Artem.Abakumov@ua.ac.be}
\author{Gustaaf Van Tendeloo}
\email{staf.vantendeloo@ua.ac.be}
\affiliation{EMAT, University of Antwerp, Groenenborgerlaan 171, B-2020 Antwerp, Belgium}
\author{Helge Rosner}
\affiliation{Max Planck Institute for Chemical Physics of Solids, N\"{o}thnitzer
Str. 40, 01187 Dresden, Germany}


\begin{abstract}
We report on the crystal structure of the quantum magnet (CuCl)LaNb$_2$O$_7$ that was controversially described with respect to its structural organization and magnetic behavior. Using high-resolution synchrotron powder x-ray diffraction, electron diffraction, transmission electron microscopy, and band structure calculations, we solve the room-temperature structure of this compound [$\alpha$-(CuCl)LaNb$_2$O$_7$] and find two high-temperature polymorphs. The $\gamma$-(CuCl)LaNb$_2$O$_7$ phase, stable above 640~K, is tetragonal with $a_{\sub}=3.889$~\r A, $c_{\sub}=11.738$~\r A, and the space group $P4/mmm$. In the $\gamma$-(CuCl)LaNb$_2$O$_7$ structure, the Cu and Cl atoms are randomly displaced from the special positions along the $\{100\}$ directions. The $\beta$-phase ($a_{\sub}\times 2a_{\sub}\times c_{\sub}$, space group $Pbmm$) and the $\alpha$-phase ($2a_{\sub}\times 2a_{\sub}\times c_{\sub}$, space group $Pbam$) are stable between 640~K and 500~K and below 500~K, respectively. The structural changes at 500 and 640~K are identified as order-disorder phase transitions. The displacement of the Cl atoms is frozen upon the $\gamma\rightarrow\beta$ transformation, while a cooperative tilting of the NbO$_6$ octahedra in the $\alpha$-phase further eliminates the disorder of the Cu atoms. The low-temperature $\alpha$-(CuCl)LaNb$_2$O$_7$ structure thus combines the two types of the atomic displacements that interfere due to the bonding between the Cu atoms and the apical oxygens of the NbO$_6$ octahedra. The precise structural information resolves the controversy between the previous computation-based models and provides the long-sought input for understanding (CuCl)LaNb$_2$O$_7$ and related compounds with unusual magnetic properties.
\end{abstract}

\maketitle

\section{Introduction}
\label{introduction}
Superexchange interactions underlie the magnetic phenomena in most transition metal compounds. Such interactions involve orbitals of non-magnetic cations and are therefore highly sensitive to subtle changes in the crystal structure. This high sensitivity can be used to effectively vary the properties of the system by applying external pressure\cite{waki2007} or by performing chemical substitutions.\cite{tsujimoto2010} On the other hand, it underlies the crucial importance of a precise structure determination. Inaccurate structural information can lead to irrelevant or even wrong conclusions regarding the magnetic properties of the system.\cite{janson2008,schmitt2009}

\begin{figure}
\includegraphics{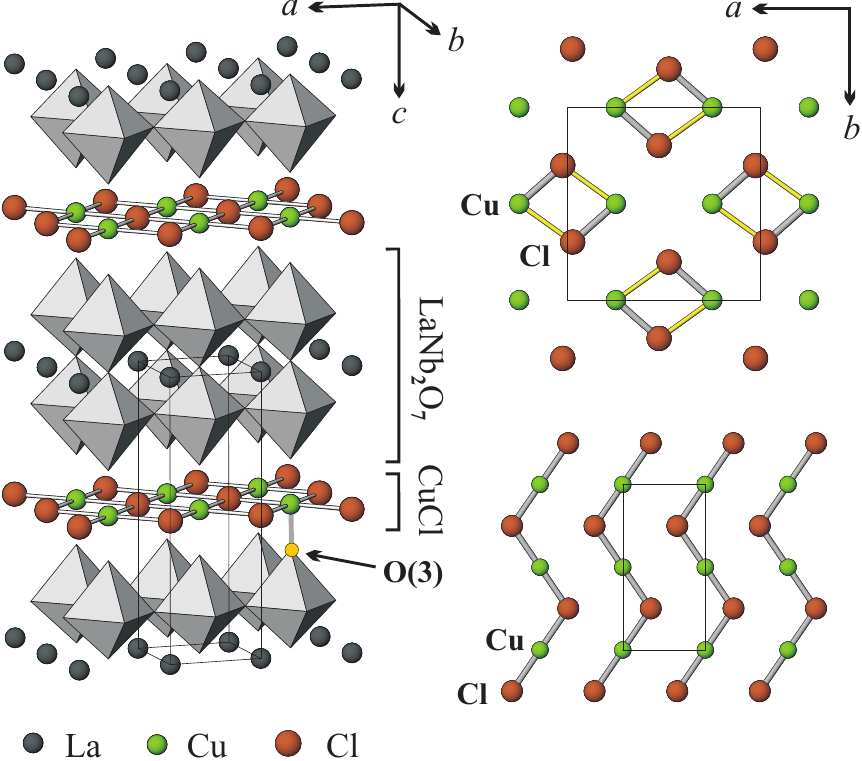}
\caption{\label{structure}
(Color online) Left panel: crystal structure of (CuCl)LaNb$_2$O$_7$. The Cu atoms are bonded to the Cl atoms and to the O(3) atoms of the NbO$_6$ octahedra. Right panel: the ordering models for the (CuCl) layers, as proposed in Refs.~\onlinecite{yoshida2007} (top) and~\onlinecite{tsirlin2009} (bottom). Thick lines show short Cu--Cl bonds. The unit cell in the left panel denotes the tetragonal subcell, according to Ref.~\onlinecite{koden1999}.
}
\end{figure}

(CuCl)LaNb$_2$O$_7$ is a parent compound for the family of layered materials, derived from perovskite-type Dion-Jacobson phases.\cite{koden1999,koden2001} In such compounds, the two-dimensional (2D) perovskite blocks alternate with (MX) layers, where M$^{2+}$ is a transition metal and X is Cl or Br (Fig.~\ref{structure}). The Cu-containing compounds appear to be especially interesting with respect to the unusual magnetic properties because of their low, spin-$\frac12$ magnetic moment that leads to quantum behavior at low temperatures.\cite{stone2006,takigawa2010} (CuCl)LaNb$_2$O$_7$ shows a spin gap and Bose-Einstein condensation of magnons in high magnetic fields.\cite{kageyama2005,kageyama2005-2,kitada2007} The bromine analog (CuBr)LaNb$_2$O$_7$ undergoes stripe antiferromagnetic (AFM) ordering,\cite{oba2006} while a related (CuBr)Sr$_2$Nb$_3$O$_{10}$ compound shows a magnetization plateau at one-third of the saturation.\cite{tsujimoto2007} All these interesting quantum effects remain poorly understood from the theoretical side due to a lack of adequate spin models that can be only developed, based on accurate structural information.

The original study proposed (CuCl)LaNb$_2$O$_7$ to have tetragonal symmetry with both Cu and Cl atoms on the four-fold axes.\cite{koden1999} However, the high atomic displacement parameter of Cl ($U_{\iso}=0.13$~\r A$^2$) evidenced a shift of the Cl atom from this high-symmetry position. If the overall tetragonal symmetry is preserved, the Cl atoms have to be disordered over several equivalent sites. On the other hand, an ordered arrangement of the Cl atoms would imply the formation of a superstructure with reduced symmetry and, presumably, with an extended unit cell. Since the orbitals of the Cl atoms mediate superexchange interactions, the precise determination of the Cl position is crucial to understand the magnetic properties.

Later studies included neutron powder diffraction\cite{caruntu2002} and synchrotron x-ray powder diffraction experiments\cite{oba2007} but neither of both showed any superstructure reflections that could arise from an ordered arrangement of the displaced Cl atoms. The superstructure reflections were observed by electron diffraction,\cite{yoshida2007} yet no quantitative structural information has been extracted from these data. Nuclear magnetic/quadrupole resonance (NMR/NQR) studies evidenced unique positions of the Cu, Cl, and La atoms and confirmed the formation of an ordered structure.\cite{yoshida2007} The subsequent analysis of electric field gradients (EFGs) showed sizable violation of the tetragonal symmetry. Based on these results, Yoshida \textit{et al.}\cite{yoshida2007} empirically established a structural model. In their model, the neighboring Cu atoms are linked into a dimer via two Cl atoms with shorter Cu--Cl bonds. The neighboring dimers are connected by longer Cu--Cl bonds only (the top right panel of Fig.~\ref{structure}). The assignment was solely based on the type of the superstructure and on the hyperfine field at the Cl site, while no quantitative evaluation of EFGs or exchange couplings was performed.

In an earlier study, we attempted an \textit{ab initio} computational approach to the problem.\cite{tsirlin2009} Employing density functional theory (DFT) calculations, we relaxed the crystal structure of (CuCl)LaNb$_2$O$_7$ and found that the system tends to undergo a Jahn-Teller-type distortion. The displacements of the Cl atoms lead to two short and two long Cu--Cl bonds. Additionally, each Cu atom has two short Cu--O bonds to the oxygens of the NbO$_6$ octahedra. Then, the short Cu--Cl bonds align along one of the intralayer directions, and chains of corner-sharing CuO$_2$Cl$_2$ plaquettes are formed (bottom right panel of Fig.~\ref{structure}). This DFT-based model readily reproduced the experimental EFGs for the Cu and Cl sites. It also explained the low energy scale of the exchange couplings and the non-trivial exchange coupling between fourth-neighbors, evidenced by the inelastic neutron scattering.\cite{kageyama2005,kageyama2005-2} Still, this computation-based model did not yield the full solution of the problem, because it did not explain the large experimental asymmetry of the EFG at the La site. Moreover, the subtle balance of ferromagnetic and antiferromagnetic superexchange couplings was evidenced by experimental studies,\cite{kageyama2005} but could not be fully reproduced in band structure calculations. Further on, a subsequent computational study\cite{ren} challenged our model and claimed a different structure that is similar to the earlier proposal by Yoshida \textit{et al.}\cite{yoshida2007}

Band structure calculations provide an effective \textit{ab initio} route of structure determination. However, they experience conceptual difficulties in the treatment of electronic correlations, relevant for many transition metal compounds. The choice of the unit cell and the crystal symmetry can also be a subtle issue. Therefore, the experimental input is still desirable to derive an unambiguous structural model. In the following, we show that high-resolution synchrotron powder x-ray diffraction (SXRD), combined with electron microscopy and assisted by band structure calculations, leads to an unambiguous structure solution and resolves the existing controversy. Moreover, the variable-temperature SXRD experiments reveal three polymorphs of (CuCl)LaNb$_2$O$_7$. These polymorphs are further denoted as $\alpha$, $\beta$, and $\gamma$ according to the temperature ranges of their stability. 

The paper is organized as follows. In Sec.~\ref{methods}, we review the experimental and computational methods. In Sec.~\ref{results}, we determine the room-temperature (RT) $\alpha$-(CuCl)LaNb$_2$O$_7$ structure and investigate the temperature evolution of the crystal structure. Then, we corroborate our conclusions by electron diffraction, electron microscopy, \textit{ab initio} structure relaxation, and EFG calculation. In Sec.~\ref{discussion}, we perform the symmetry analysis, discuss the interplay between the structures of different polymorphs, and the implications of our findings for the magnetic properties of (CuCl)LaNb$_2$O$_7$.

\section{Methods}
\label{methods}
Powder samples of (CuCl)LaNb$_2$O$_7$ were obtained via a two-step procedure. First, we prepared RbLaNb$_2$O$_7$ by firing a mixture of La$_2$O$_3$ and Nb$_2$O$_5$ with a 25\%-excess of Rb$_2$CO$_3$ in air at 1050~\Cels\ for 24 h. The resulting white powder was washed with water to obtain single-phase RbLaNb$_2$O$_7$. At the second step, RbLaNb$_2$O$_7$ was mixed with a two-fold excess of anyhydrous CuCl$_2$. To avoid any traces of moisture, the commercial anhydrous CuCl$_2$ was dried overnight at 80~\Cels\ and further handled in a glove box. The mixture of RbLaNb$_2$O$_7$ and CuCl$_2$ was pressed into a pellet, sealed into an evacuated quartz tube, and fired at 400~\Cels\ for 48~h. The resulting samples were washed with water to remove CuCl$_2$ and RbCl. Finally, the samples were dried overnight at 80~\Cels. Note that we used a higher annealing temperature compared to the previous studies: 400~\Cels\ (673~K) vs. 325~\Cels\ (598~K) in Refs.~\onlinecite{koden1999} and~\onlinecite{kageyama2005}. The higher annealing temperature does not lead to a decomposition of (CuCl)LaNb$_2$O$_7$ and rather improves the sample quality, because the high-temperature $\gamma$-modification of (CuCl)LaNb$_2$O$_7$ is formed during the annealing and transformed into the RT $\alpha$-modification upon cooling (see Sec.~\ref{evolution}). Our samples revealed sharp superstructure reflections at RT, while the samples prepared at 325~\Cels\ showed a more diffuse scattering at certain positions. According to XRD, the samples were single-phase and fully reproducible. Several samples were additionally checked with magnetization measurements (MPMS SQUID, $2-380$~K temperature range, $0.1-5.0$~T field range) and showed essentially identical susceptibility curves that were also in agreement with previous reports.\cite{kageyama2005}

To test the sample purity, we collected laboratory powder x-ray diffraction data using a Huber G670f Guinier camera (CuK$_{\alpha1}$ radiation, Image Plate detector). The high-resolution SXRD patterns were measured at the ID31 beamline of the European Synchtrotron Radiation Facility (ESRF) with a constant wavelength of about 0.4~\r A. The data were collected by eight scintillation detectors, each preceded by a Si (111) analyzer crystal, in the angle range $2\theta=1-40$~deg. The powder sample was contained in a thin-walled borosilicate glass capillary with an external diameter of 0.5~mm. To achieve good statistics and to avoid the effects of the preferred orientation, the capillary was spun during the experiment. The sample was cooled below RT in a He-flow cryostat (temperature range $100-300$~K) or heated above RT with a hot-air blower ($300-660$~K). 

The structure refinement was performed in the \texttt{JANA2000} program.\cite{jana2000} The conventional profile matching and the Le Bail fitting procedure were strongly impeded by the weak reflection splitting due to a small orthorhombic distortion. Therefore, we also used the Rietveld refinement to extract the unit cell parameters at different temperatures. To achieve maximum accuracy, the zero shift and the asymmetry parameter were fixed according to the refinement of the RT data. The symmetry analysis was done in the \texttt{ISODISPLACE} program.\cite{isodisplace}

The samples for transmission electron microscopy (TEM) were prepared by crushing the powder sample in ethanol and depositing it on a holey carbon grid. Selected area electron diffraction (ED) patterns were recorded using a Philips CM20 microscope. High-angle annular dark-field scanning TEM (HAADF STEM) images were taken with a Tecnai G2 microscope. High-resolution TEM (HRTEM) images were recorded on JEOL 4000EX and Tecnai G2 microscopes. The HRTEM images were simulated by means of the \texttt{JEMS} software. 

To confirm the refined crystal structure, we performed \textit{ab initio} geometry relaxation and evaluated the EFGs for the $\alpha$-modification, the only ordered polymorph of (CuCl)LaNb$_2$O$_7$. The band structure calculations were done with the FPLO9.00-33 code\cite{fplo} that performs full-potential DFT calculations applying the basis set of atomic-like local orbitals. Both the local density approximation (LDA)\cite{pw92} and the generalized gradient approximation (GGA)\cite{pbe} for the exchange-correlation potential were used. To account for correlation effects, we used the mean-field LSDA+$U$/GGA+$U$ approaches with the around-mean-field double-counting correction scheme.\cite{amf} The on-site Coulomb repulsion and the on-site exchange parameters were fixed at $U_d=5.5$~eV and $J_d=1$~eV, according to the previous studies.\cite{tsirlin2009} The variation in $U_d$ in the range of $5.5\pm 1$~eV and the type of the magnetic ordering had little influence on the resulting geometry and on the EFG values. Residual forces in the relaxed structures did not exceed 0.01~eV/\r A. All the calculations were performed in the 48-atom unit cell with lattice parameters derived from the experimental structure refinement. The $k$-mesh included 192 points in the symmetry-irreducible part of the first Brillouin zone. The general information on the band structure of (CuCl)LaNb$_2$O$_7$ has been reported in a previous publication.\cite{tsirlin2009}

\section{Results}
\label{results}
\subsection{Room-temperature\\ $\alpha$-(CuCl)LaNb$_2$O$_7$ crystal structure}
Contrary to the previous reports,\cite{koden1999,oba2007} the laboratory XRD powder pattern shows several weak reflections, violating the earlier proposed tetragonal unit cell of (CuCl)LaNb$_2$O$_7$ with $a_{\sub}=b_{\sub}\simeq 3.87$~\r A and $c_{\sub}\simeq 11.8$~\r A. The high-resolution SXRD experiment reveals 23 additional reflections of this type. The resulting pattern can be indexed in a supercell with $a\simeq b\simeq 2a_{\sub}=7.74$~\r A and $c\simeq 11.8$~\r A. The temperature evolution of the superlattice reflections (see Sec.~\ref{evolution} and Fig.~\ref{patterns}) rules out their possible extrinsic origin and assigns them to the (CuCl)LaNb$_2$O$_7$ phase. 

\begin{figure*}[!ht]
\includegraphics[scale=0.95]{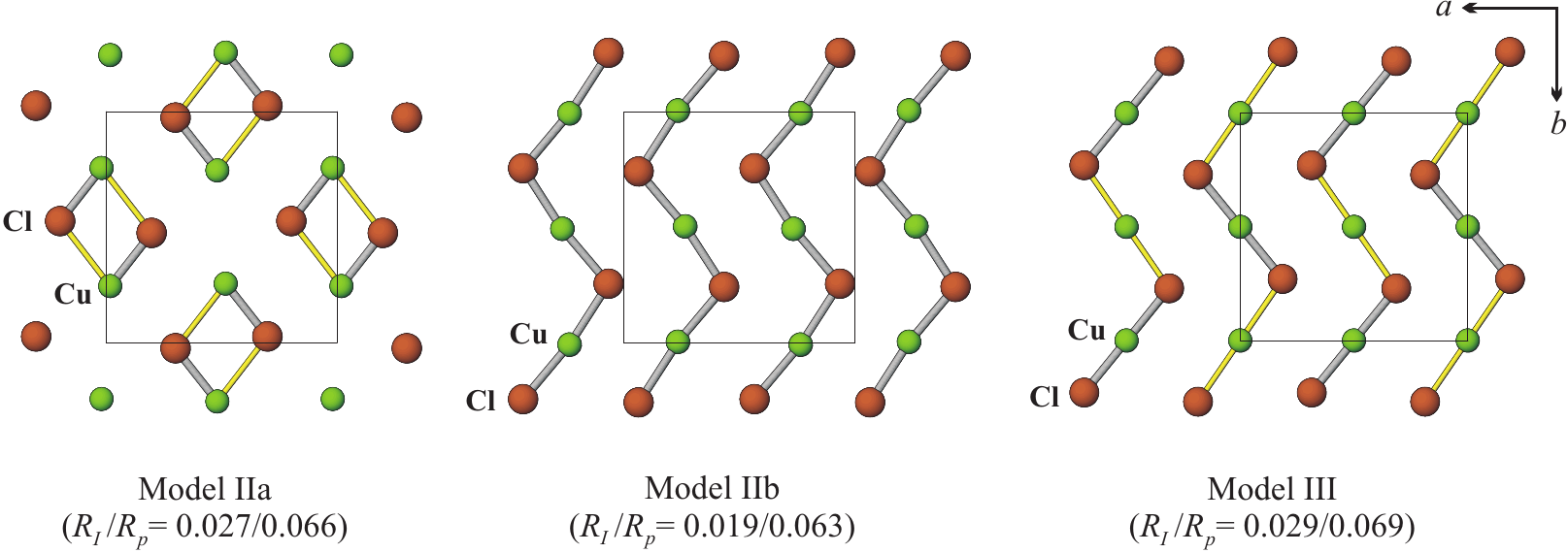}
\caption{\label{models}
(Color online) Three models of the $z=1/2$ (CuCl) layer based on the refinement of the RT SXRD data. Gray (thick) and yellow (thin) lines show the Cu--Cl bonds in the ranges $2.3-2.5$~\r A and $2.5-2.8$~\r A, respectively. The numbers under the plots list the refinement residuals $R_I/R_p$.
}
\end{figure*}
The room-temperature SXRD data are consistent with the tetragonal symmetry and do not demonstrate any apparent reflection splitting that could originate from the symmetry reduction. The ED patterns, however, clearly indicate an orthorhombic symmetry (see Sec.~\ref{electron}). Moreover, the temperature evolution of the SXRD pattern also evidences a subtle difference between the $a$ and $b$ parameters (see Fig.~\ref{parameters}) and suggests an orthorhombic symmetry. At RT, the difference between $a$ and $b$ is, by chance, negligible. Therefore, no reflection splittings could be observed within the present experimental resolution.\cite{note1} As one takes into account the actual orthorhombic symmetry, the space group $Pbam$ or its subgroup $Pba2$ are identified, based on the reflection conditions $k=2n$ for $0kl$ and $h=2n$ for $h0l$. These reflection conditions were also confirmed by the ED study (see Sec.~\ref{electron}). In the following, we adopt the $Pbam$ space group for the structure refinement. We did not find any signature of the disorder that could point to an acentric crystal  structure. 

In (CuCl)LaNb$_2$O$_7$, a superstructure can be caused by cooperative tilts of the NbO$_6$ octahedra and/or by atomic displacements in the (CuCl) layers. The possible modes of the tilting distortion and the atomic displacements are classified by considering a virtual transformation of the parent $a_{\sub}\times a_{\sub}\times c_{\sub}$ tetragonal structure with the $P4/mmm$ space symmetry to the orthorhombic $2a_{\sub}\times 2b_{\sub}\times c_{\sub}$ superstructure with the $Pbam$ space symmetry. To denote the tilting patterns, we use the Glazer's notation\cite{glazer} in the form $a^-b^0c^+$, where $a$, $b$, and $c$ stand for different tilt angles around the respective crystal axes, $+/-$ denote the in-phase/out-of-phase rotations, respectively, and $0$ indicates the absence of the tilt. The atomic displacements within the (CuCl) layers are confined to the $ab$ plane due to the mirror plane symmetry of the $Pbam$ space group. If all the atoms with a certain $y$ coordinate have the same shifts along $a$ (or, similarly, all the atoms with a certain $x$ coordinate have the same shifts along $b$), the resulting displacement is called ``in-phase''. In contrast, the ``out-of-phase'' displacement implies the shifts of these atoms in opposite directions.

Since different combinations of the atomic displacements and octahedral tilts are possible, we construct several starting models for the refinement. These models are conveniently classified by the position of the inversion center that determines the arrangement of the $a$ and $b$ glide planes, while these planes further confine possible distortions. If an atom lies on the glide plane, the displacements along the glide direction should be "in-phase". In contrast, the glide plane passing between the two atoms imposes the "out-of-phase" displacements along the glide direction. The $a$ and $b$ glide planes do not allow for the in-phase octahedral tilts around $a$ and $b$. The out-of-phase tilt around $a$ is only possible if the $a$ glide plane passes through the Nb atom, while the $b$ glide plane contains the common oxygen atom of the two octahedra. Any other arrangement of the glide planes would completely forbid the tilts around $a$ and $b$ ($a^0b^0$). The following starting models are possible:

I) The inversion center is at the Cl site, hence no displacements of the Cl atoms are possible. The Cu and Nb atoms lie on the glide planes. Then, the Cu atoms can shift "in-phase" along both $a$ and $b$ axes, but the octahedral tilts around $a$ and $b$ are forbidden (the tilting pattern $a^0b^0c^+$).

II) The inversion center is placed between the two Cu atoms. Then, the $a$ glide plane contains the Cu and Nb atoms and passes between the Cl atoms, while the $b$ glide plane contains the Cl atoms and passes between the Cu atoms. The opposite arrangement (Cu and Nb on $b$ planes, Cl on $a$ planes) is also possible and leads to two different options:

IIa) Left panel of Fig.~\ref{models}: the Cl atoms follow the "in-phase" displacement along $b$ and the "out-of-phase" displacement along $a$. The Cu atoms show the "in-phase" displacement along $a$ and the "out-of-phase" displacement along $b$. The tilt system is $a^-b^0c^0$.      

IIb) Middle panel of Fig.~\ref{models}: the Cl atoms follow the "in-phase" displacement along $a$ and the "out-of-phase" displacement along $b$. The Cu atoms show the "in-phase" displacement along $b$ and the "out-of-phase" displacement along $a$. The tilt system is $a^0b^-c^0$.

III) Right panel of Fig.~\ref{models}: the inversion center is at the Cu site, hence the displacements of the Cu atoms are forbidden. The Cl atoms lie on the glide planes can be displaced "in-phase" along both $a$ and $b$ axes. No octahedral tilting is possible ($a^0b^0c^0$).    

Model I can be rejected immediately: it does not allow to displace the Cl atoms, while the previous studies evidenced the pronounced shift of Cl (see Sec.~\ref{introduction}). The model III implies two inequivalent Cu positions and formally violates the NMR/NQR data that suggest a single Cu position.\cite{yoshida2007} However, the positions with similar local environment are not necessarily resolved by nuclear resonance techniques. Therefore, we also considered the model III in the structure refinement. 

The models IIa and IIb show a cooperative tilt of the NbO$_6$ octahedra around either the $a$ or $b$ axis. In the model IIa, the structure of the (CuCl) layer resembles the model by Yoshida \textit{et al.}\cite{yoshida2007} with pairs of Cu atoms connected into dimers via shorter Cu--Cl bonds (left panel of Fig.~\ref{models}). The model IIb can be derived from our computation-based model\cite{tsirlin2009} and shows chains of corner-sharing CuO$_2$Cl$_2$ plaquettes (middle panel of Fig.~\ref{models}). In the model IIa, the local environment of Cu is fairly irregular with one very short (2.26~\r A) and one longer (2.75~\r A) Cu--Cl bond. These bonds have a \emph{cis}-configuration with a Cl--Cu--Cl angle of $76^{\circ}$. In contrast, the model IIb yields a typical CuO$_2$Cl$_2$ plaquette with two Cu--Cl distances of about 2.4~\r A in a \emph{trans}-configuration (a Cl--Cu--Cl angle of $173^{\circ}$). In the model III, the two inequivalent Cu atoms  show different local environments, with one of these atoms being clearly underbonded (all the Cu--Cl distances are above 2.5~\r A). 

\begin{figure}
\includegraphics{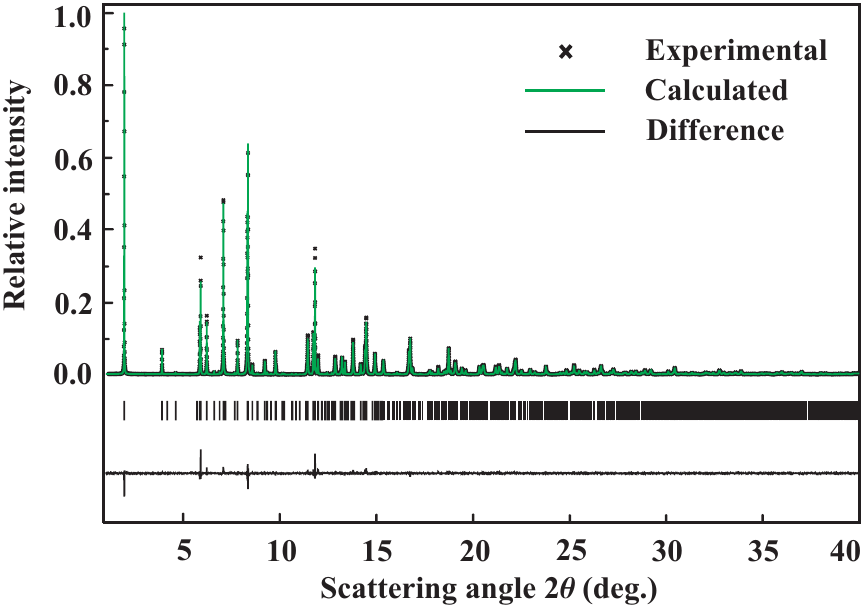}
\caption{\label{refinement}
(Color online) Experimental, calculated, and difference SXRD patterns for $\alpha$-(CuCl)LaNb$_2$O$_7$. Ticks show the reflection positions.
}
\end{figure}
The refinement residuals for the three models are summarized in Fig.~\ref{models}. Based on this comparison, we select model IIb as the correct structure of $\alpha$-(CuCl)LaNb$_2$O$_7$. This choice is further supported by the results of band structure calculations (Sec.~\ref{band}) and by the symmetry analysis (Sec.~\ref{discussion}). It is also consistent with empirical crystal chemistry arguments that suggest the CuO$_2$Cl$_2$ plaquettes as a typical local environment of Cu$^{2+}$ in oxychlorides.\cite{cu2ocl2} Atomic coordinates for the model IIb are listed in Table~\ref{coordinates}, while representative interatomic distances are given in Table~\ref{distances}. Figure~\ref{refinement} shows experimental, calculated, and difference SXRD patterns.

In the structure of $\alpha$-(CuCl)LaNb$_2$O$_7$, the NbO$_6$ octahedra are tilted around the $b$ axis, while the chains of the corner-sharing CuO$_2$Cl$_2$ plaquettes also align along the $b$ direction. The resulting crystal structure is very similar to our previous DFT-based model that, for the sake of simplicity, did not consider the tilts of the NbO$_6$ octahedra.\cite{tsirlin2009} The tilting distortion doubles the unit cell along $a$ and is coupled to the shifts of the Cu atoms. Yet the local environment of Cu and the topology of the layer are retained.

\begin{table}
\caption{\label{coordinates}
Atomic coordinates of $\alpha$-(CuCl)LaNb$_2$O$_7$ at room temperature: $a=7.76290(8)$~\r A, $b=7.76197(7)$~\r A, $c=11.73390(6)$~\r A, space group $Pbam$, and $R_I/R_p= 0.019/0.063$. Atomic displacement parameters $U_{\iso}$ for oxygen atoms were constrained.
}
\begin{ruledtabular}
\begin{tabular}{ccrrrr}
  Atom & Position & $x$ & $y$ & $z$ & $U_{\iso}$%
\footnote{The $U_{\iso}$ values are given in $10^{-2}$~\r A$^2$.} \\\hline 
  Cu   & $4h$ & 0.7343(4) & 0.5058(8)   & $\frac12$  & 0.89(4) \\
  Cl   & $4h$ & 0.5653(4) & 0.2429(10)  & $\frac12$  & 2.3(1)  \\
  La   & $4g$ & 0.0002(2) & 0.2556(2)   & 0          & 0.44(1) \\
  Nb   & $8i$ & 0.7486(2) & 0.4985(3)   & 0.80827(4) & 0.30(1) \\
  O(1) & $4f$ & 0         & $\frac12$   & 0.8201(9)  & 0.51(6) \\
  O(2) & $8i$ & 0.254(1)  & 0.749(2)    & 0.8471(8)  & 0.51(6) \\
  O(3) & $8i$ & 0.234(1)  & 0.013(2)    & 0.6569(3)  & 0.51(6) \\
  O(4) & $4g$ & 0.770(2)  & 0.487(2)    & 0          & 0.51(6) \\
  O(5) & $4e$ & 0         & 0           & 0.858(1)   & 0.51(6) \\
\end{tabular}
\end{ruledtabular}
\end{table}
\subsection{Temperature evolution of the crystal structure}
\label{evolution}
The unusual magnetic properties of (CuCl)LaNb$_2$O$_7$ are observed at low temperatures, which makes the low-temperature crystal structure most relevant for magnetic studies. On the other hand, the evolution of the superstructure reflections above RT can give important information about their origin. Therefore, we investigated (CuCl)LaNb$_2$O$_7$ both above and below RT. 

\begin{figure}
\includegraphics{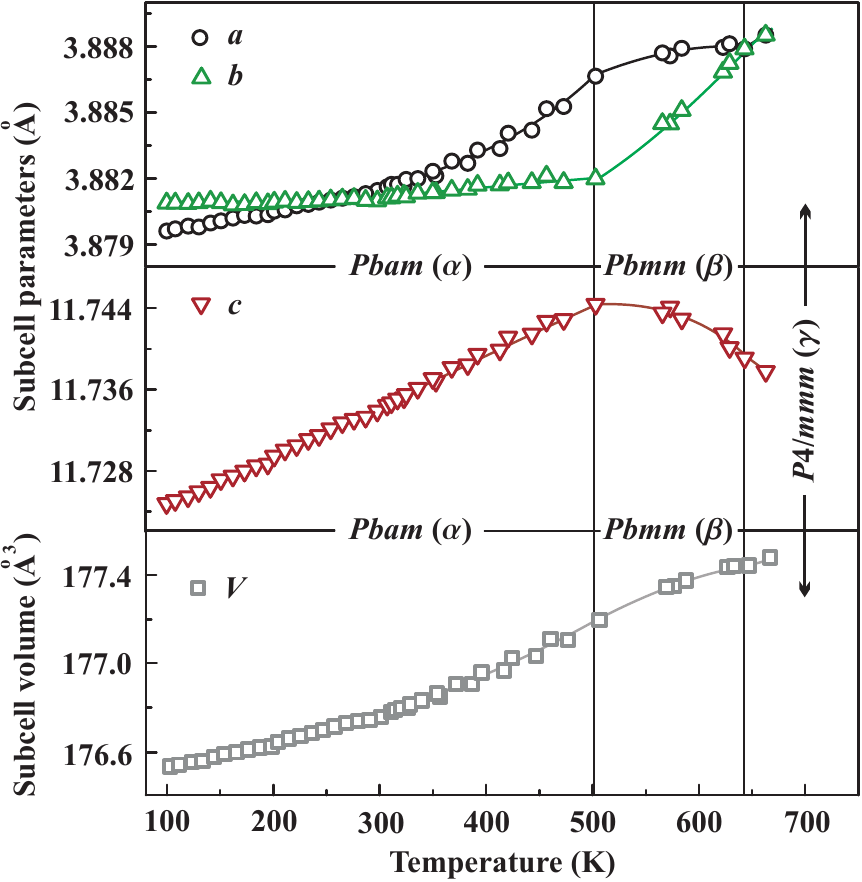}
\caption{\label{parameters}
(Color online) Temperature dependence of the subcell lattice parameters and the subcell volume for (CuCl)LaNb$_2$O$_7$. Error bars do not exceed the size of the symbols.
}
\end{figure}
\begin{figure}
\includegraphics{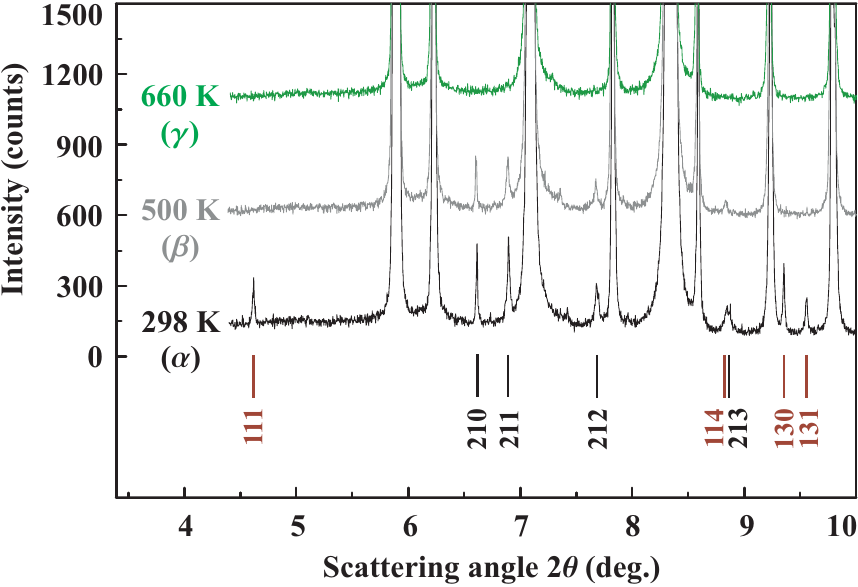}
\caption{\label{patterns}
(Color online) Temperature evolution of the XRD patterns. The reflections are indexed in the $2a_{\sub}\times 2a_{\sub}\times c_{\sub}$ supercell, only the superstructure reflections are labeled. The reflections with even $h$ and odd $k$ are attributed to the $k_1=[0,\frac12,0]$ propagation vector, while the reflections with both $h$ and $k$ odd can be assigned to the $k_2=[\frac12,0,0]$ propagation vector. The patterns are offset for clarity. The maximum intensity for the main reflections is about 50000 counts.
}
\end{figure}

Figure~\ref{parameters} shows the temperature dependence of the lattice parameters and the unit cell volume. The orthorhombic strain, defined as $\varepsilon=(a_{\sub}-b_{\sub})/(a_{\sub}+b_{\sub})$, is negative at 100~K and increases upon heating. The strain amounts to zero at $\simeq 260$~K and reaches its maximum value at 500~K. There is no indication of a structural transformation in the $100-500$~K temperature range, thus the RT $\alpha$-(CuCl)LaNb$_2$O$_7$ crystal structure is a reliable model for the investigation of the low-temperature properties. At 500~K, the orthorhombic strain starts to decrease, because of the $\alpha\rightarrow\beta$ phase transition between the two orthorhombic polymorphs. The strain vanishes at 640~K indicating a $\beta\rightarrow\gamma$ orthorhombic-to-tetragonal phase transition. Although the two phase transitions are clearly visible in the SXRD data, we failed to observe them with differential scanning calorimetry (DSC), presumably, due to the low entropy change associated with these transitions.\cite{note2} The phase transitions are accompanied by a change in the intensities of the superstructure reflections (Fig.~\ref{patterns}). Above 640~K, the SXRD pattern of the $\gamma$-(CuCl)LaNb$_2$O$_7$ phase was indexed in a simple $a_{\sub}\times a_{\sub}\times c_{\sub}$ tetragonal unit cell. No reflection conditions were identified, suggesting the $P4/mmm$ space group for the $\gamma$-phase. The SXRD pattern of the $\beta$-(CuCl)LaNb$_2$O$_7$ phase shows additional reflections corresponding to the $k_1=[0,\frac12,0]$ propagation vector. The resulting $a_{\sub}\times 2a_{\sub}\times c_{\sub}$ unit cell is orthorhombic, and the space symmetry is $Pbmm$ ($0kl$, $k=2n$ reflection conditions).\cite{note3} The second set of superstructure reflections corresponds to the $k_2=[\frac12,0,0]$ propagation vector\cite{note4} and appears below 500~K at the $\beta\rightarrow\alpha$ phase transition.  

\begin{figure*}[!ht]
\includegraphics{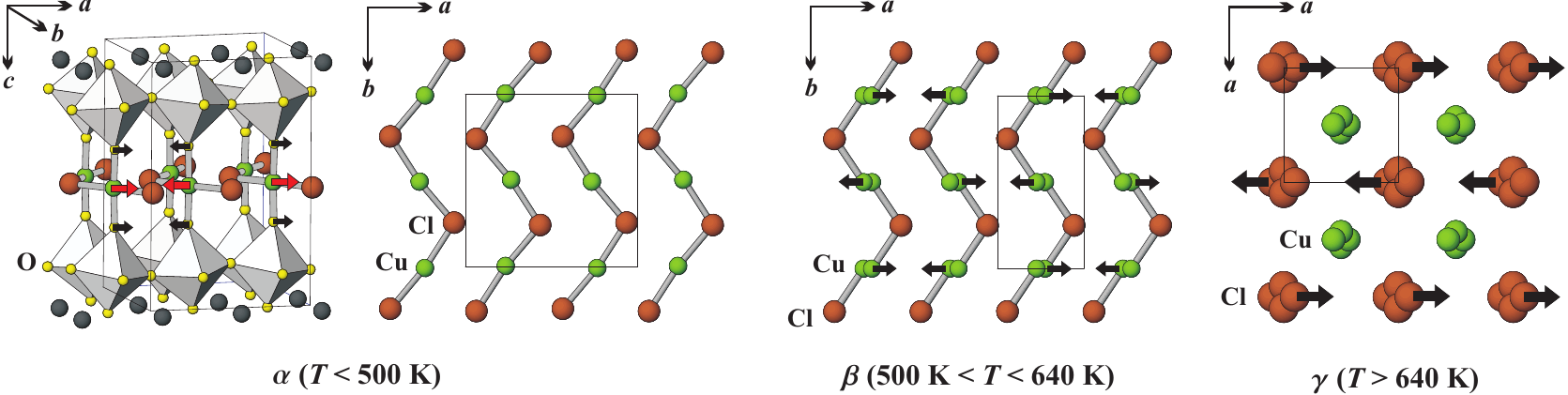}
\caption{\label{modifications}
(Color online) The structure of (CuCl)LaNb$_2$O$_7$ at different temperatures. Left panel: an overall view with the tilted NbO$_6$ octahedra and the fully ordered [CuCl] layer below 500~K ($\alpha$-modification). Middle panel: the [CuCl] layer with the disordered Cu atoms (between 500~K and 640~K, $\beta$-modification). Right panel: the disordered structure of the [CuCl] layer (above 640~K, $\gamma$-modification). The arrows show the primary atomic displacements. Note the interconnected displacements of Cu and O in the left panel.
}
\end{figure*}

The continuous change in the unit cell volume (Fig.~\ref{parameters}) suggests that both phase transitions are of the second order. The $P4/mmm\rightarrow Pbmm\rightarrow Pbam$ symmetry evolution is consistent with this assumption (see Sec.~\ref{discussion}). The octahedral tilting distortion is completely suppressed in $\beta$-(CuCl)LaNb$_2$O$_7$, and the superstructure is solely caused by the atomic displacements in the (CuCl) layer. We find ordered, "in-phase" shifts of the Cl atoms along the $a$ direction (see the middle panel of Fig.~\ref{modifications}). The Cu atom can be placed into the special $2d$ position $(\frac12,0,\frac12)$, but the shift of Cu to a split, four-fold position reduced the refinement residuals and the atomic displacement parameter of Cu. Therefore, the split position of Cu should be preferred. The refined atomic positions are listed in Table~\ref{coordinates-beta}. The $\beta$-(CuCl)LaNb$_2$O$_7$ structure essentially matches our DFT-based model\cite{tsirlin2009} with respect to the unit cell and the crystal symmetry. The only difference is the partial disorder due to the split position of the Cu atom. In this position, each Cu atom has a short Cu--Cl bond of 2.34~\r A and a longer bond of 2.57~\r A. Similar to the $\alpha$-(CuCl)LaNb$_2$O$_7$ structure, these bonds show the \emph{trans}-configuration with a Cl--Cu--Cl angle close to $180^{\circ}$.

\begin{table}
\caption{\label{coordinates-beta}
Atomic coordinates of $\beta$-(CuCl)LaNb$_2$O$_7$ at 500~K: $a=3.88666(3)$~\r A, $b=7.76396(5)$~\r A, $c=11.74441(7)$~\r A, space group $Pbmm$, and $R_I/R_p=0.024/0.062$. Atomic displacement parameters $U_{\iso}$ for oxygen atoms were constrained.
}
\begin{ruledtabular}
\begin{tabular}{ccrrrr}
  Atom & Position & $x$ & $y$ & $z$ & $U_{\iso}$%
\footnote{The $U_{\iso}$ values are given in $10^{-2}$~\r A$^2$} \\\hline 
   Cu\footnote{Occupancy factor $g=\frac12$}
       & $4j$ & 0.5482(9)  & 0          & $\frac12$  & 1.13(6) \\
   Cl  & $2f$ & 0.1145(9)  & $\frac34$  & $\frac12$  & 3.7(1)  \\
   La  & $2e$ & 0.0001(3)  & $\frac34$  & 0          & 0.81(1) \\
   Nb  & $4h$ & $\frac12$  & 0          & 0.80757(4) & 0.54(1) \\
  O(1) & $4g$ & 0          & 0          & 0.8515(7)  & 1.29(6) \\
  O(2) & $4k$ & 0.508(2)   & $\frac14$  & 0.8343(7)  & 1.29(6) \\
  O(3) & $4h$ & $\frac12$  & $\frac12$  & 0.6567(3)  & 1.29(6) \\
  O(4) & $2c$ & $\frac12$  & 0          & 0          & 1.29(6) \\
\end{tabular}
\end{ruledtabular}
\end{table}
\begin{table}
\caption{\label{coordinates-gamma}
Atomic coordinates of $\gamma$-(CuCl)LaNb$_2$O$_7$ at 660~K: $a=3.88852(2)$~\r A, $c=11.73778(7)$~\r A, space group $P4/mmm$, and $R_I/R_p=0.027/0.066$. Atomic displacement parameters $U_{\iso}$ for oxygen atoms were constrained.
}
\begin{ruledtabular}
\begin{tabular}{ccrrrr}
  Atom & Position & $x$ & $y$ & $z$ & $U_{\iso}$%
\footnote{The $U_{\iso}$ values are given in $10^{-2}$~\r A$^2$} \\\hline 
   Cu\footnote{Occupancy factor $g=\frac14$}
       & $4o$ & 0.5662(8) & $\frac12$  & $\frac12$  & 1.24(7) \\
   Cl\footnote{Occupancy factor $g=\frac14$}
       & $4m$ & 0.096(2)  & 0          & $\frac12$  & 4.7(2)  \\
   La  & $1a$ & 0         & 0          & 0          & 1.08(2) \\
   Nb  & $2h$ & $\frac12$ & $\frac12$  & 0.19305(4) & 0.69(1) \\
  O(1) & $4i$ & 0         & $\frac12$  & 0.1582(2)  & 1.88(6) \\
  O(2) & $2h$ & $\frac12$ & $\frac12$  & 0.3436(3)  & 1.88(6) \\
  O(3) & $1c$ & $\frac12$ & $\frac12$  & 0          & 1.88(6) \\
\end{tabular}
\end{ruledtabular}
\end{table}
In the tetragonal $\gamma$-(CuCl)LaNb$_2$O$_7$ structure, neither Cu nor Cl atoms lie on the four-fold axes. Our structure refinement (Table~\ref{coordinates-gamma} and the right panel of Fig.~\ref{modifications}) shows that both Cu and Cl partially occupy the split four-fold positions with a shortest Cu--Cl distance of 2.30~\r A. This implies that copper prefers to keep short bonds to the chlorine atoms, but the spatial arrangement of these short bonds is now fully disordered.

\subsection{Electron diffraction and transmission electron microscopy}
\label{electron}
\begin{figure}
\includegraphics[scale=1.1]{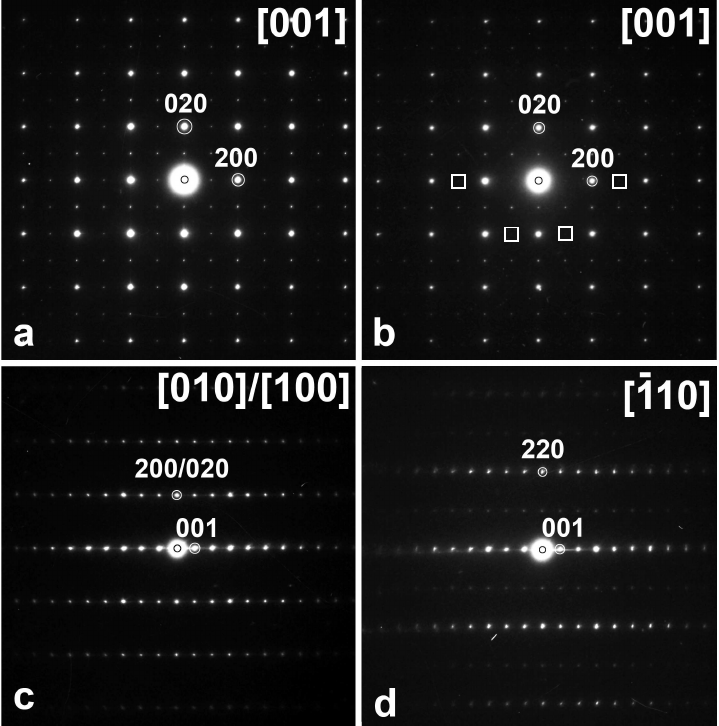}
\caption{\label{electron diffraction}
Electron diffraction patterns of $\alpha$-(CuCl)LaNb$_2$O$_7$. The superstructure reflections are present in the patterns $a$, $b$, and $d$. The pattern $a$ shows the pseudotetragonal symmetry due to the $90^{\circ}$ rotational twin domains, while the pattern $b$ is taken with a smaller aperture and points to the orthorhombic symmetry. White boxes denote several positions of the missing reflections. The patterns $c$ and $d$ mainly show the subcell reflections.
}
\end{figure}

Electron diffraction patterns of $\alpha$-(CuCl)LaNb$_2$O$_7$ (Fig.~\ref{electron diffraction}) were readily indexed in the pseudo-tetragonal unit cell with $a\simeq b\simeq 7.75$~\r A and $c\simeq 11.8$~\r A, as proposed by the SXRD data. These patterns clearly evidence the superstructure in the $ab$ plane. The $[001]$ ED pattern in Fig.~\ref{electron diffraction}a has apparent tetragonal symmetry. However, a closer inspection of different regions of the crystals with a smaller selected area aperture revealed the areas where the intensity distribution contradicts the four-fold symmetry (Fig.~\ref{electron diffraction}b). The pseudotetragonal $[001]$ ED pattern (Fig.~\ref{electron diffraction}a) is then a sum of the ED patterns from 90$^{\circ}$ rotational twin domains with orthorhombic symmetry. Due to the small difference between the $a$ and $b$ lattice parameters and due to the twinning, the $[010]$ and $[100]$ ED patterns could not be distinguished (Fig.~\ref{electron diffraction}c). The reflection conditions $h=2n$ for $h0l$ and $k=2n$ for $0kl$, derived from the ED patterns, are in agreement with the $Pbam$ space group. In the $[001]$ ED pattern, the forbidden $0k0$, $k\neq 2n$ reflections appear due to the multiple diffraction. 

We found that the ordering in the (CuCl)LaNb$_2$O$_7$ structure is very sensitive to electron-beam irradiation. The superstructure reflections with $h+k\neq 2n$ disappear rapidly and irreversibly under an intense electron beam, while the reflections with $h+k=2n$ remain. HRTEM images demonstrate that these changes are related to the amorphization of the (CuCl) layers under the electron beam, whereas the LaNb$_2$O$_7$ blocks are less affected. The evolution of the structure under the electron-beam irradiation is surprisingly different from the effect of heating (Sec.~\ref{evolution}). The heating first removes the tilts of the NbO$_6$ octahedra, while the (partial) ordering in the (CuCl) layers survives up to 640~K. In contrast, the electron beam ``melts'' the (CuCl) layers and weakly changes the perovskite blocks. Weak diffuse intensity lines along $c^*$ are visible on the $[010]/[100]$ and $[\bar110]$ ED patterns (Fig.~\ref{electron diffraction}c and d) and can be attributed to a local irradiation damage of the (CuCl) layers. 

\begin{figure}
\includegraphics[scale=0.95]{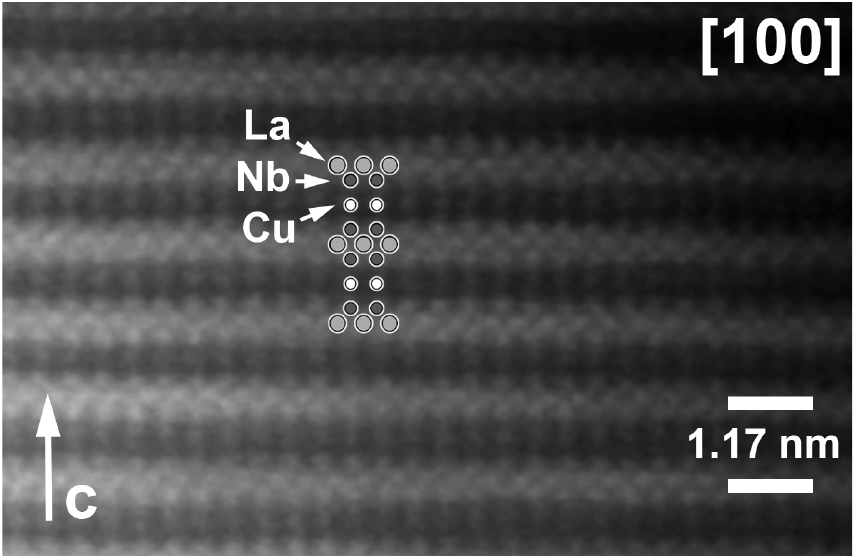}
\caption{\label{haadf}
$[100]$ HAADF STEM image of $\alpha$-(CuCl)LaNb$_2$O$_7$. The cation positions are overlaid.
}
\end{figure}
\begin{figure}
\includegraphics[scale=0.95]{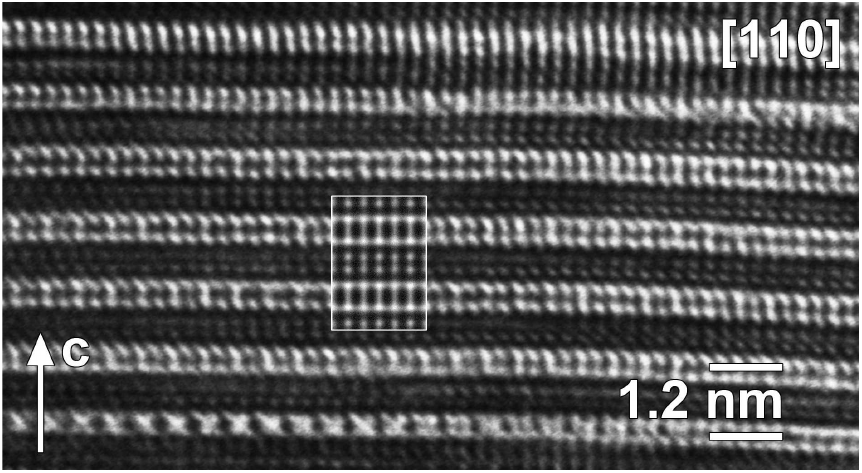}
\caption{\label{110 hrem}
$[110]$ HRTEM image of $\alpha$-(CuCl)LaNb$_2$O$_7$. The simulated image ($f=31$~nm and $t=3.3$~nm) is shown as an inset and outlined with a white rectangle.
}
\end{figure}

Remarkably, the HAADF STEM mode does not introduce such a severe beam damage as the HRTEM imaging. On the HAADF STEM image, the intensity is related to the average atomic number of an atomic column as $I\approx{Z^n}$ $(n=1-2)$. The $[100]$ HAADF STEM image of (CuCl)LaNb$_2$O$_7$ (Fig.~\ref{haadf}) reveals a defectless sequence of the cationic layers along the $c$ axis. However, the oxygen and chlorine atoms have too low atomic numbers to be observed in this imaging mode. The oxygen and chlorine displacements can be revealed by the HRTEM technique. In order to reduce the beam damage, the HRTEM images were collected under the minimum possible beam intensity and keeping the exposure time as short as possible. On the $[110]$ HRTEM image (Fig.~\ref{110 hrem}), the rows of brighter dots correspond to the empty spacings in the layers of apical oxygen atoms of the NbO$_6$ octahedra. The dots within the brighter rows are arranged into alternating pairs with different brightness due to the displacements of the apical oxygen atoms upon the octahedral tilt. The theoretical HRTEM image (defocus $f=31$~nm, sample thickness $t=3.3$~nm) calculated with the $Pbam$ model IIb is in agreement with the contrast of the experimental image. Note, however, that the displacements in the (CuCl) layers can not be observed in this structure projection. 

\begin{figure}
\includegraphics[scale=0.95]{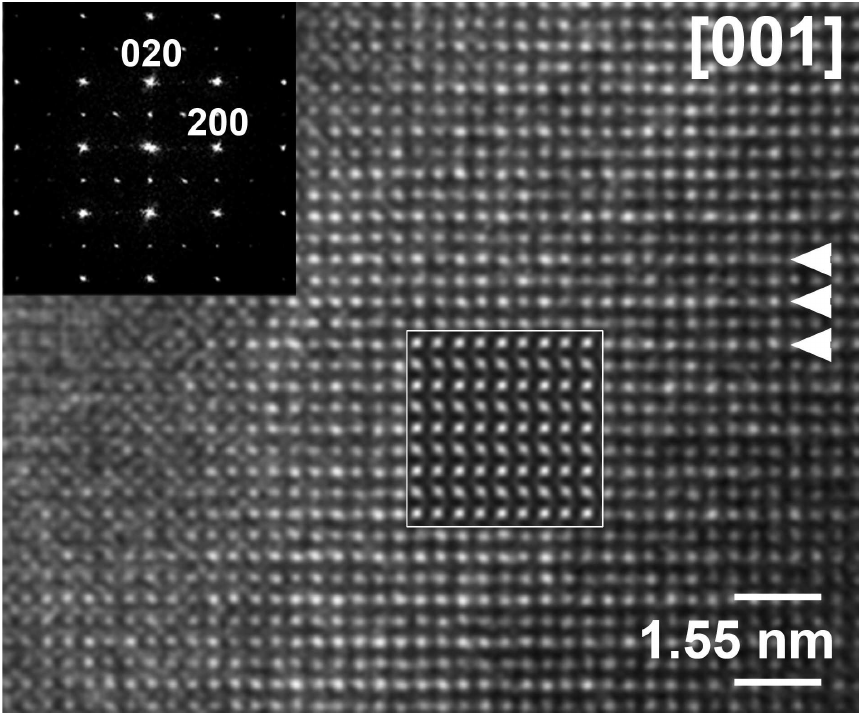}
\caption{\label{001 hrem}
$[001]$ HRTEM image of $\alpha$-(CuCl)LaNb$_2$O$_7$. The Fourier transformed image is shown at the top left corner. The simulated image ($f=-38$~nm and $t=3.5$~nm) is shown as an inset and outlined with a white rectangle. Arrowheads mark the rows of brighter dots.
}
\end{figure}

The atomic displacements in the (CuCl) layer are visible when the structure is viewed along the $[001]$ axis (Fig.~\ref{001 hrem}). Under these imaging conditions, the brighter dots correspond to the projections of the LaCl columns. The  superstructure is clearly visible on this image as an alternation of brighter and less bright rows (some of the brighter rows are marked with arrows in Fig.~\ref{001 hrem}). The presence of this superstructure is also confirmed by the Fourier transformed image, which shows the spots corresponding to the doubling of the lattice parameters in the \textit{ab} plane. The simulated HRTEM image ($f=-38$~nm, $t=3.5$~nm) calculated with the $Pbam$ model IIb is in agreement with the experimental data.

\subsection{Structure relaxation and electric field gradients}
\label{band}
The experimental information on the unit cell and the crystal symmetry imposes helpful constraints for the computational analysis. The calculations, on the other hand, can evaluate the relative stability of different structural models. If all the atomic coordinates are fixed to the experimental positions, model IIb shows the lowest energy. Models IIa and III lie higher in energy by 0.53~eV/f.u. and 0.81~eV/f.u., respectively. As the atomic positions are relaxed, models IIa and IIb converge to the same solution that closely resembles model IIb with short Cu--Cl bonds in the \emph{trans}-configuration. The relaxation of model III yields a similar structure with slightly higher energy, because the Cu atoms are constrained to the two-fold positions. Overall, the comparison of total energies and the \textit{ab initio} structure relaxation strongly support model IIb and discard the other two models of the $\alpha$-(CuCl)LaNb$_2$O$_7$ structure.

\begin{table}
\caption{\label{distances}
Interatomic distances (in~\r A) and angles (in~deg.) in $\alpha$-(CuCl)LaNb$_2$O$_7$, as found from the structure refinement and from the LSDA+$U$ or GGA+$U$ \textit{ab initio} relaxations.
}
\begin{ruledtabular}
\begin{tabular}{cccc}
         & Refinement        & LSDA+$U$       & GGA+$U$        \\\hline
  Cu--Cl & 2.41(1)           & 2.32           & 2.38           \\
         & 2.43(1)           & 2.32           & 2.38           \\
Cu--O(3) & $2\times 1.85(1)$ & $2\times 1.85$ & $2\times 1.85$ \\
  Cu--Cl--Cu & 107.1(2)      & 113.7          & 109.7          \\\hline
  Nb--O(1) & 1.96(1)         & 1.97           & 1.93           \\
  Nb--O(2) & 1.97(2)         & 1.99           & 2.00           \\
           & 2.02(2)         & 2.00           & 2.00           \\
  Nb--O(3) & 1.78(1)         & 1.79           & 1.79           \\
  Nb--O(4) & 2.26(1)         & 2.26           & 2.29           \\
  Nb--O(5) & 2.02(1)         & 2.03           & 2.04           \\
\end{tabular}
\end{ruledtabular}
\end{table}

It is also instructive to compare the details of the experimental and relaxed geometries. Table~\ref{distances} shows the interatomic distances and the Cu--Cl--Cu angle, as found from the structure refinement and from the LSDA+$U$/GGA+$U$ relaxations. The LSDA+$U$ relaxation underestimates the Cu--Cl distances, while GGA+$U$ shows better agreement with the experiment. Thus, GGA+$U$ is the best available computational approach to model the crystal structure of (CuCl)LaNb$_2$O$_7$. Our refined interatomic distances are in remarkable agreement with the previous structure refinement based on the neutron diffraction data.\cite{caruntu2002} The neutron experiment did not show any superstructure reflections, and the ordered arrangement of the Cl atoms could not be found. However, the Cl atom was moved to the four-fold position with an occupancy of $\frac14$, yielding a Cu--Cl distance of 2.40~\r A ($2.41-2.43$~\r A in our refinement). According to Ref.~\onlinecite{caruntu2002}, the Cu--O distance is 1.84~\r A; this also matches our findings (1.85~\r A). The reference to the neutron data ensures the remarkable accuracy of the obtained structural information, despite the huge difference in the scattering power of oxygen and the heavy (La, Nb) atoms.

Structural distortions can be traced by EFGs that have been determined in a NMR experiment.\cite{yoshida2007} Table~\ref{efg} lists EFGs calculated for three different models of the RT structure: $V_{zz}$ is the leading principal component of the tensor, while $\eta=(V_{yy}-V_{xx})/V_{zz}$ is the asymmetry. The results for the Cu and Cl sites are similar to our previous report: the $z$ axis of the EFG tensor lies in the $ab$ plane for Cu and aligns along $c$ for Cl.\cite{tsirlin2009} Note however that the model IIa leads to a pronounced asymmetry of the Cu EFG ($\eta=0.43$) due to the very irregular local environment of the Cu atom. The model IIb shows a lower asymmetry ($\eta=0.04$) in agreement with the experimental result ($\eta=0.10$). Finally, the model III yields fairly different EFGs (well above the experimental resolution) for the two inequivalent Cu sites and is readily discarded.

\begin{table}
\caption{\label{efg}
Calculated EFG's for three models of the $\alpha$-(CuCl)LaNb$_2$O$_7$ structure. $V_{zz}$ is the principal component of the tensor (in $10^{21}$~V/m$^2$), $\eta$ is the tensor asymmetry. Model III has two inequivalent Cu positions. The experimental values are taken from Ref.~\onlinecite{yoshida2007} and do not allow to determine the sign of $V_{zz}$. The structural models are shown in Fig.~\ref{models}.
}
\begin{ruledtabular}
\begin{tabular}{crrrrrr}
  & & \multicolumn{1}{l}{Cu} & \multicolumn{2}{c}{Cl} & \multicolumn{2}{c}{La} \\
             & $V_{zz}$    & $\eta$    & $V_{zz}$ & $\eta$ & $V_{zz}$ & $\eta$ \\\hline
  Model IIa  & $-12.9$     & 0.43      & $-16.6$  & 0.73   & $-17.3$  & 0.08   \\
  Model IIb  & $-11.4$     & 0.04      & $-18.3$  & 0.50   & $-14.2$  & 0.57   \\
  Model III  & \multicolumn{1}{c}{$-9.2$/$-12.2$} & \multicolumn{1}{c}{0.16/0.22} & $-18.0$  & 0.44   & $-21.2$  & 0.26   \\
  Experiment & 11.6        & 0.10      & $-14.2$  & 0.56   & 6.5      & 0.70   \\
\end{tabular}
\end{ruledtabular}
\end{table}
The tilts of the NbO$_6$ octahedra allow to reproduce the sizable asymmetry of the EFG at the La site. The calculated asymmetry for the model IIb ($\eta=0.57$) is in agreement with the experimental value ($\eta=0.70$). In contrast, the model IIa yields a very low asymmetry and does not fit to the experimental findings. Compared to the experimental data, the $V_{zz}$ value remains rather high in all the three models. This discrepancy can be caused by $4f$ orbitals of La or $4d$  orbitals of Nb that lie close to the Fermi level and are subject to correlation effects. Our present approach accounts for the correlations in the Cu $3d$ shell only and treats all the other orbitals on the LDA level.  To achieve better agreement with the experiment, more sophisticated computational techniques are necessary. Apart from the deviation for $V_{zz}$, the calculated EFG's strongly support the model IIb and clearly disfavor the two other models.

The \emph{ab initio} structure relaxation also provides insight into the interplay of the $k_1$ and $k_2$ superstructures in $\alpha$-(CuCl)LaNb$_2$O$_7$. To reveal the influence of the tilting distortion on the structure of the (CuCl) layers, we performed a relaxation with fixed positions of La, Nb, and O. We used the regular structure of the (LaNb$_2$O$_7$) blocks, as found from the refinement at 660~K. Taking the Cu position as $(\frac34+x,\frac12+y,\frac12)$, we find a negligible displacement of the Cu atoms ($x=-0.0001$ and $y=0.0001$) for the relaxed structure with the fixed (LaNb$_2$O$_7$) block, compared to $x=-0.0157$ and $y=0.0058$ for the experimental crystal structure. Thus, the tilting distortion tends to shift the Cu atoms from their ``parent'' ($\frac34,\frac12,\frac12$) position. In the $\alpha$-(CuCl)LaNb$_2$O$_7$ structure, the octahedra tilt around the $b$ axis, hence the axial O(3) atoms (see Fig.~\ref{structure}) displace along the $a$ axis. The Cu atoms follow these displacements and also shift along $a$ (left panel of Fig.~\ref{modifications}). The interconnected displacements of the Cu and O(3) atoms underlie the coupling between the $k_1$ and $k_2$ superstructures in (CuCl)LaNb$_2$O$_7$. This coupling can be understood as the tendency to keep the regular four-fold coordination of Cu with Cu--O bonds perpendicular to the Cu--Cl bonds. The preference to the regular coordination is also evidenced by the \emph{trans}-arrangement of the Cu--Cl bonds and rules out alternative structural models that would suggest a highly irregular local environment of the Cu atom.

\section{Discussion}
\label{discussion}
In the preceding section, we have derived the $\alpha$-(CuCl)LaNb$_2$O$_7$ structure by considering different ordering patterns in the (CuCl) layer (Fig.~\ref{models}) and selected the model IIb, based on the comparison of the refinement residuals, the total energies, and the EFGs. Alternatively, one can consider the $\alpha$-(CuCl)LaNb$_2$O$_7$ structure as the derivative of the high-temperature tetragonal $\gamma$-(CuCl)LaNb$_2$O$_7$ phase. Assuming second-order phase transitions, the structural distortion in the low-temperature polymorph should follow one of the irreducible representations for the space group of the high-temperature polymorph. 

The $Pbmm$ $\beta$-structure can be derived from the high-temperature $P4/mmm$ $\gamma$-structure by the $X_3^-$ irreducible representation [here $X$ denotes the $(0,\frac12,0)$ point in the reciprocal space and implies a doubling of the $b$ parameter]. This irreducible representation decouples the four Cl atoms within one position and splits the four equivalent Cu atoms into two pairs. The leading distortion is the ``in-phase'' shift of the Cl atoms. The Cu atoms are disordered over two equivalent positions around $(\frac12,0,\frac12)$. 

The $\beta\rightarrow\alpha$ transition follows the $Z_4^+$ irreducible representation of the $Pbmm$ space group, where $Z$ is $(\frac12,0,0)$ and corresponds to a doubling of the $a$ parameter [$Z$ is $(0,0,\frac12)$ in the standard $Pmma$ setting]. The Cu atoms shift along $a$ in the "out-of-phase" manner (the middle panel of Fig.~\ref{modifications}). The O(3) atoms also shift along $a$, leading to the $a^0b^-c^0$ octahedral tilting pattern. The structural changes at 500~K and 640~K can be identified as atomic displacements order-disorder phase transitions. The evolution of the crystal symmetry is in agreement with the second-order nature of these transitions.  Note that the intermediate $Pbmm$ structure allows for the ``in-phase'' shifts of the Cl atoms only, thus imposing the similar arrangement of the Cl atoms in the $\alpha$-structure. The alternative structural model by Yoshida \textit{et al.}\cite{yoshida2007} (model IIa) would require a dramatic rearrangement of the (CuCl) layers, which is unlikely. 

Since the experimentally determined structure of the (CuCl) layer in the model IIb is rather similar to our previous DFT-based model,\cite{tsirlin2009} certain conclusions on the electronic structure and magnetic properties can be derived. The regular (tetragonal) structure of (CuCl)LaNb$_2$O$_7$ leads to the orbital degeneracy of Cu. Atomic displacements in the (CuCl) layers are primarily caused by the tendency to lift the orbital degeneracy. The resulting coordination of the Cu atoms is the CuO$_2$Cl$_2$ plaquette with the magnetic (half-filled) $d_{x^2-y^2}$ orbital lying in the plane of the plaquette. The respective bands are well separated from other Cu $3d$ states and ensure the one-orbital scenario, typical for cuprates. 

The lobes of the magnetic $d_{x^2-y^2}$ orbital point toward Cl and O atoms. The strong $\sigma$-overlap between Cu $3d$ and Cl $3p$ orbitals leads to a rather unusual superexchange scenario with strong coupling between fourth neighbors (Cu--Cu distance of about 8.7~\r A) in agreement with the inelastic neutron scattering results.\cite{kageyama2005} A quantitative evaluation of this unusual long-range coupling depending on the Cl position has been given in Ref.~\onlinecite{tsirlin2009}. 

The strong hybridization of the magnetic Cu $3d_{x^2-y^2}$ orbital with the oxygen orbitals leads to a somewhat three-dimensional nature of the system. In Ref.~\onlinecite{tsirlin2009}, we have shown that this hybridization along with the low-lying $4d$ states of Nb results in a sizable interlayer coupling of about 15~K, compared to $50-60$~K for the couplings in the $ab$ plane. The strong hyperfine coupling at the Nb site\cite{yoshida2007} can be taken as an experimental signature of the strong Cu--O--Nb hybridization. 

Although the reference to the DFT-based model clarifies the electronic structure of the compound, we should preclude the reader from the straight-forward transfer of the microscopic magnetic model. Superexchange couplings are highly sensitive to fine details of the crystal structure. Therefore, sizable changes in the exchange couplings are expected. According to Table~\ref{distances}, the LSDA+$U$ structure relaxation leads to a Cu--Cl--Cu angle of $113.8^{\circ}$, while the experimental value is about $107.1^{\circ}$. The angles at the ligand atoms are known to have a dramatic influence on the magnitude and even on the sign of the superexchange couplings. Following simple microscopic arguments in the spirit of Goodenough-Kanamori rules,\cite{goodenough} we suggest that the AFM nearest-neighbor interaction along the $b$ direction should be reduced compared to our previous calculations\cite{tsirlin2009} or even changed towards a FM interaction. Since this interaction is the leading coupling in the original spin model, substantial changes in the resulting magnetic ground state should be expected. The shifts of the Cu atoms result in a large number of inequivalent Cu--Cu distances and, consequently, in a large number of different exchange interactions. The detailed analysis of the spin model is a challenging task involving extensive band structure calculations, model simulations, and quantitative comparison to the experimental data. Such a study is presently underway.\cite{model}

Using the combination of different techniques, the present study gives a detailed insight into the chemical modifications of (CuCl)LaNb$_2$O$_7$. We show that the atomic displacements in this crystal structure are interconnected, and a tilting distortion of the perovskite block induces atomic displacements in the (CuCl) layers (see the left panel of Fig.~\ref{modifications}). This explains why the substitutions in the perovskite block (e.g., Ta for Nb) have pronounced effects on the magnetism,\cite{kitada2009} despite the leading interactions are expected within the (CuCl) layers for the whole compound family. The interplay of the atomic displacements also provides a key to the puzzling behavior of (CuBr)Sr$_2$Nb$_3$O$_{10}$ and related compounds where tiny changes in the perovskite block destroy the peculiar $\frac13$-magnetization plateau.\cite{tsujimoto2007,tsujimoto2008} 

Based on the temperature evolution of the (CuCl)LaNb$_2$O$_7$ structure, we are able to propose a general mechanism for structural distortions in the family of Cu-based quantum magnets as well as in isostructural (MCl)LaNb$_2$O$_7$ compounds (M = Cr, Mn, Fe, Co).\cite{viciu2002,viciu2003,viciu2003-2} The regular (tetragonal) structure of the [CuX] (X = Cl or Br) layer is highly unfavorable due to the orbital degeneracy of Cu (Ref.~\onlinecite{tsirlin2009}) and unsuitable Cu--X distances. The strain is released by the displacements of the Cu and X atoms leading to shorter and longer Cu--X bonds. Below certain temperature [640~K in (CuCl)LaNb$_2$O$_7$], such shifts become cooperative, but the Cu atoms remain disordered due to their bonding to the oxygen atoms (middle panel of Fig.~\ref{modifications}). The tilting distortion allows to shift the Cu atoms and results in their ordered arrangement (left panel of Fig.~\ref{modifications}). Thus, the low-temperature structure is fully ordered and can be reliably determined using conventional diffraction techniques. The interplay of the atomic displacements plays a crucial role in this compound family. The specific tilting distortion sets an ordered configuration of the Cu atoms and has a strong effect on the magnetic properties. Further studies of related materials should challenge this mechanism and disclose the influence of different atoms (Cl vs. Br and Nb vs. Ta) on the transition temperatures and distortion patterns.

In summary, we have determined the room-temperature crystal structure of (CuCl)LaNb$_2$O$_7$ and observed two structural order-disorder phase transitions in this compound. The room-temperature structure combines the tilting distortion of the (LaNb$_2$O$_7$) perovskite blocks with ordered displacements of the Cu and Cl atoms within the (CuCl) layers. The tilting distortion disappears above 500~K and leads to the partially disordered arrangement of the Cu atoms while a further transition at 640~K destroys the ordering in the (CuCl) layers completely. The room-temperature structure shows a clear preference of Cu to the four-fold square-like coordination with two O and two Cl neighbors in a CuO$_2$Cl$_2$ plaquette. The plaquettes share corners and form chains running along the $b$ direction. This type of the structure organization is basically consistent with our previous computation-based structural model and resolves the controversy between several competing proposals. The obtained structural information is the necessary and long-sought input for understanding (CuCl)LaNb$_2$O$_7$ and related compounds with unusual magnetic properties.

\textit{Note added:} after the submission of the original manuscript, the results of Ref.~\onlinecite{ren} were updated.\cite{cheng-2} The reported space group and the DFT-based structural model are in agreement with our \emph{experimental} structure solution, although the
interatomic distances and angles are slightly different, similar to Table~IV. Ren and Cheng\cite{ren,cheng-2} accidentally refer their model to the earlier proposal by Yoshida \textit{et al.}, (Ref.~\onlinecite{yoshida2007}) despite the corner-sharing arrangement of the CuO$_2$Cl$_2$ plaquettes essentially matches our original model (Ref.~\onlinecite{tsirlin2009}) that was also based on DFT. Additionally, Tassel \textit{et al.}\cite{tassel} refined the structure of $\alpha$-(CuCl)LaNb$_2$O$_7$ using neutron diffraction data and also arrived to the orthorhombic $Pbam$ symmetry. The resulting atomic positions are in close agreement with our results.

\acknowledgments
We are grateful to ESRF for providing the beam time at ID31 and particularly acknowledge Andrew Fitch for his faultless help during the data collection. We also acknowledge Yurii Prots and Horst Borrmann for laboratory XRD measurements, Katrin Koch and Klaus Koepernik for implementing EFGs in the FPLO code, Stefan Hoffmann for the DSC measurements, as well as Dirk Wulferding and Peter Lemmens for disclosing the Raman spectra of (CuCl)LaNb$_2$O$_7$. A.T. was funded by Alexander von Humboldt foundation.


\begin{thebibliography}{36}
\bibitem [{not(){\natexlab{a}}}]{note1}%
  \BibitemOpen
  \bibinfo {note} {Apart from the small difference between $a$ and $b$ at RT,
  the splittings are also concealed by a strong overlap of the reflections
  above $2\theta=15^{\circ}$.}\BibitemShut {Stop}%
\bibitem [{not(){\natexlab{b}}}]{note2}%
  \BibitemOpen
  \bibinfo {note} {The DSC measurement was performed with a Perkin Elmer DSC
  8500 instrument in the temperature range $80-1000$~K in argon atmosphere with
  a heating rate of 2~K/min. The powder sample was placed in a corundum
  crucible. The only anomaly was observed above 850~K and could be assigned to
  the decomposition of (CuCl)LaNb$_2$O$_7$.}\BibitemShut {Stop}%
\bibitem [{not(){\natexlab{d}}}]{note4}%
  \BibitemOpen
  \bibinfo {note} {We imply the propagation vector for the
  $\beta$-modification, thus only the $a$ parameter is doubled.}\BibitemShut
  {Stop}%
\end{thebibliography}
\end{document}